\documentclass[runningheads]{llncs}
\usepackage[utf8]{inputenc}
\usepackage{times}
\usepackage{url}
\usepackage{graphicx,subfig}
\usepackage{comment}
\usepackage{amssymb}

\begin{document}
\sloppy

% that's a working title, MN
\title{Blockguard: Adaptive Blockchain Security}
% \titlerunning{Adaptive Blockchain Security}

\author{Shishir Rai, Kendric Hood, Mikhail Nesterenko, and Gokarna Sharma}

\institute{Department of Computer Science, Kent State University, Kent, OH 44242, USA \\
\email{srai@kent.edu, khood5@kent.edu, mikhail@cs.kent.edu, sharma@cs.kent.edu}}
\maketitle

\begin{abstract}
We consider the problem of varying the security of blockchain transactions according to their importance. This adaptive security is achieved by using variable size consensus committees. To improve performance, such committees function concurrently. We present two  algorithms that allow adaptive security by forming concurrent variable size consensus committees on demand. One is based on a single joint blockchain, the other is based on separate sharded blockchains.
For in-committee consensus, our algorithms may use various available byzantine-robust fault tolerant algorithms (BFT). We implement synchronous BFT, asynchronous BFT and proof-of-work consensus. We thoroughly evaluate the performance of our adaptive security algorithms. 
% add the result of the work, MN

\end{abstract}
%\keywords{blockchain, consensus, security, distributed algorithms.}

\thispagestyle{plain} % forces page numbering
\pagestyle{plain}

% call it "algorithm", not "protocol", 
% no "blocks" we call them "transactions"
% do not use "system", use "network"
% do not use nodes, processors, processes, etc. only "peers"
% independent ledgers are not to be called blockchain
%https://github.com/khood5/distributed-consensus-abstract-simulator.git

\section{Introduction}
\noindent\textbf{Blockchain.} A secure distributed ledger, \emph{blockchain} allows a decentralized network of peers to register a sequence of transactions despite potentially malicious actions of a minority of peers. Blockchain technology is poised to revolutionize a variety of fields: from currency and payment systems that are impervious to state and corporate manipulation~\cite{nakamoto,ethereum}, to automatically enforced contracts~\cite{Abdellatif2018,ConcurrencySmartContracts}, to internet-of-things massive data recording~\cite{iotchain}.

Peers acting maliciously are \emph{Byzantine}~\cite{byzantine}: a Byzantine peer may exhibit arbitrary behavior. Such malicious peers are controlled by \emph{adversary} that uses them to compromise the blockchain. To overcome the adversary, peers coordinate their collective decision.  This decision may be achieved using classic byzantine-robust coordinated consensus~\cite{pbft,paxos,byzantine} or novel proof-of-work~\cite{nakamoto,ethereum}. 

In a coordinated consensus algorithm, the peers exchange messages to arrive at a uniform conclusion. Such an algorithm usually requires every peer to know the identities of all other peers. That is, a completely connected network topology is necessary. A coordinated consensus algorithm operates correctly so long as the number of Byzantine peers is less than its resiliency threshold. This threshold is a fraction of the committee size.  In a fast changing peer-to-peer network, such strict membership requirement is problematic. 
%Moreover, the adversary may launch a Sybil attack~\cite{sybil}, creating a large number of fake identities to defeat the network. To control the network membership, a new peer needs to obtain permission to join the network. That is, the system are \emph{permissioned}.
% introduce synchronous vs. asynchronous consensus
In a proof-of-work algorithm, the peers compete to solve a computationally intensive problem. All peers accept the solution of the peer who solves the problem first. Such algorithms do not require membership or network topology maintenance.  The algorithm operates correctly provided that the computation power of the Byzantine peers does not exceed the power of the correct peers. That is, the resiliency threshold of a proof-of-work algorithm is expressed in terms of the computation power.
However, such an algorithm  needs extensive computing resources and has poorer throughput and latency. % need citations 
This limits the scalability of proof-of-work networks.

%
% talk about proof-of-stake, Ourboros~\cite{ouroboros}
%
 
There are plenty of studies enhancing the scalability of proof-of-work networks~\cite{solida,decker2016bitcoin,bitcoinng,algorand,byzcoin,honeybadger,thunderella,phantomGhostTAG} , optimizing membership maintenance of the coordinated consensus schemes or joining the two approaches~\cite{hybrid}. 
Ultimately, the scalability of both coordinated consensus and proof-of-work networks is limited by the need to broadcast the information to all peers. To avoid this broadcast, the network needs to be logically split up or sharded. In \emph{sharding}, the network is separated into independent committees that process transactions concurrently. Sharded committees may use either coordinated or competitive consensus.

%
% One promising approach is to use a history of previous blockchain participation for committee selection. For example, select the peers that published to the blockchain most frequently or most recently~\cite{}. However, the committee thus selected has to process all submitted transactions. Such single committee may become a bottleneck and limit the scalability of the algorithm.

Sharding may potentially resolve performance issues. However, the separate committees do not provide the same protection as the complete network. Indeed, to compromise transaction processing, the adversary needs to exceed the resiliency threshold of the consensus algorithm for the single committee. Sharding also introduces the problem of cross-shard transactions where transaction inputs and outputs span shards~\cite{crossChain,omniledger,elastico,rapidchain}.

\ \\
\textbf{Adaptive security.} Thus, performance and security are cross-purposes of blockchain design. We propose to mitigate this problem with \emph{adaptive security}. Specifically, the client may request the security level for the submitted transaction on the basis of its importance. For example, the security level of purchasing a cup of coffee, may be lower than that of buying a car. The fees for registering a higher security level transaction in the blockchain may be higher. The greater security level is achieved by composing a larger committee which is less vulnerable to adversarial attack since it has a higher resiliency threshold. 

A stream of various security level transactions requires an \emph{adaptive security algorithm} that provides an appropriate size committee for each transaction. A naive solution would sequentially process transactions at the highest security level. However, employing an excessively large committee would waste network resources while sequential transaction processing may result in low throughput and high transaction waiting time. Instead, it is more efficient to assemble the consensus committees on demand and allow concurrency in transaction processing.

In this paper, we propose two such security algorithms: Composite Blockguard and Dynamic Blockguard. Composite Blockguard assembles a committee from groups of peers that maintain independent ledgers. Dynamic Blockguard selects a committee from the processes that most recently wrote to the shared ledger. Both algorithms may operate with many consensus algorithms. 

\ \\
\textbf{Our contribution.} We state the problem of adaptive security and propose two efficient algorithms that solve it. We evaluate their performance with major consensus algorithms: PBFT, SBFT and proof-of-work. We measure their throughput, transaction waiting time and resistance to Byzantine peer corruption. Our results suggest that the adaptive security provides an effective trade-off between network performance and security without significant increase in network complexity or major architectural modifications. Thus, it should be adopted by current blockchain networks.

\ \\
\textbf{Related Work.}
%\section{Related Work}
In a classic Byzantine consensus algorithm, a leader is elected and replaced if it is found faulty. The algorithm by Castro and Liskov~\cite{pbft}, known as PBFT, requires weak network synchronization for leader change and tolerates up to $f < n/3$ Byzantine peers. Synchronous network provides more information. A peer can determine whether its neighbor is faulty if the message is not received. An efficient synchronous algorithm by Abraham et al.~\cite{sbft}, that we call SBFT, tolerates up to $f < n/2$ faults and achieves consensus in 4 expected rounds of message exchanges.
Honeybadger~\cite{honeybadger} and Thunderella~\cite{thunderella} are newer approaches to cooperative consensus that try to improve its performance.

% Byteball, not citing
% Algorand
% The Tangle, Iota

%
%
% basic consensus algorithms
%
% PoW, SBFT, PBFT, HoneyBadger
%

%Thunderella~\cite{thunderella} strives to achieve higher confirmation rate by using an accelerator, also called a leader. This  accelerator proposes transitions to be added to the blockchain. Each peer must then collect signatures from other peers. Each peer will also save it's current state to another Proof-of-Work blockchain used as a log of events in the system. If a transaction is submitted but not purposed by the accelerator within a safety parameter, then Thunderella will fall back to a PoW consensus to select a new accelerator. 

%in the case where leader is honest. If the leader is corrupt, the algorithm falls back to proof-of-work consensus. 

%
%
% sharding
%
There is a number of blockchain sharding algorithms~\cite{rscoin,byzcoin,omniledger,elastico,rapidchain}.
% Elastico
% Algorand needs to be explained before OmniLedger
% OmniLedger 2nd uses Elastico, ByzCoin, and RSCoin
% BizCoin 1.5
% RapidChain 3rd optimizes OmniLedger
RSCoin~\cite{rscoin} uses central bank authority to regulate data and sharding distribution while using a peer-to-peer network for transaction registration. In Elastico~\cite{elastico}, only the agreement algorithm is sharded, the shared blockchain is replicated at every peer.
% Peers compute a proof-of-work problem to determine which committee to join.
Each committee runs PBFT. The agreed value is sent to the reference committee. This reference committee then broadcasts this value to the whole network. The cross-shard transactions have to lock multiple committees. %Algorand\cite{algorand}
OmniLedger~\cite{omniledger} shards both data and agreement portions of the network % system?
thus improving on the scalability of Elastico. It improves the cross-transaction locking mechanism and eliminates committee assignment security issues present in Elastico. RapidChain~\cite{rapidchain} further improves the sharded design. RapidChain optimizes overall message communication. It improves on the  cross-shard transaction locking mechanism of Elastico and OmniLedger by moving the locking of input transactions from the client to the committee. RapidChain also improves on cross committee communication by introducing a committee routing mechanism and optimizes peer churn handling.  None of the above sharding algorithms consider adaptive security.

% need a bias-resistant random generation protocol for leader election (algorand), Elastico has a biased random
% generation as explained in Omniledger

%
% mention DAGs
%

\section{Definitions and Consensus Algorithms}
\subsection{Definitions}
A set of $n$ \emph{peer processes} (or \emph{peers}) forms a network to maintain the blockchain. 
The \emph{blockchain} is a sequence of blocks or transactions. We use the terms interchangeably, i.e. we assume that a block contains a single transaction. A \emph{transaction} is a unit of blockchain recording. Each subsequent transaction is cryptographically linked to the previous one. The first transaction in the blockchain is the \emph{genesis} transaction. 

% no blocks! we call them transactions henceforward.

Each transaction has a unique identifier. The payload (content) of a transaction is immaterial. Any peer may generate a new transaction. Such peer is \emph{generating}. Peers do not share memory. Peer communication is through messages.
One peer may communicate with any other peer. This communication ability is always bi-directional.  A peer \emph{broadcasts} a message if it sends it to all other peers.
%The connected peers are \emph{neighbors}.
Message delivery is FIFO. There is no message loss.  Messages cannot be forged. Specifically, every peer signs its message and all other peers have ways of verifying this signature.

Peers are either \emph{honest} or \emph{Byzantine}. 
A set of peers that cooperate to approve a transaction despite actions of  Byzantine peers is a \emph{consensus committee}.
% There is no more than a fixed $f$ Byzantine peers in a committee.
% explain what "confirms" means

\ \\
\textbf{Adversary}. All Byzantine peers behave as if controlled by a single \emph{adversary} aiming to cause maximum amount of damage to the network.  The consensus algorithm may discover the activity of a Byzantine peer, detect its identity and exclude it from further operation. However, the adversary may adapt by compromising honest peers. 
So as not to exceed the maximum number of allowed Byzantine peers $f$, once the adversary claims another Byzantine peer, it needs to allow one of the already corrupted peers to become honest. This is \emph{peer shuffle}. We assume that the peer shuffle happens only when peers are idle.   That is, during consensus, a peer is either honest or Byzantine. 
% no DoS attack!

\ \\
\textbf{Sharding.} A  \emph{(recording) group} is a set of processes that maintain a single blockchain. There are as many groups as there are separate blockchains. In case of sharding, a peer in the consensus committee that approves a certain transaction in a blockchain does not necessarily belong to the group that records it. However, a peer may belong to only one recording group and only one consensus committee at a time.

\subsection{Consensus Algorithms}
\textbf{PBFT.} The committee of peers elect the \emph{leader}.
% leader id may change depending on the round, may need to say it, KH
The leader is unambiguously determined by the identities of the peers of the committee.
% request phase
A peer that generates a transaction sends it to the leader. 
% pre-prepare phase
The leader runs consensus on every arriving transaction consecutively. Once a transaction reaches the leader, it sends a \emph{pre-prepare} message with this transaction to all the committee peers.
% prepare phase
After a non-leader peer receives the pre-prepare message, the peer broadcasts a prepare message to the committee. 
% commit phase
Once a peer receives $2f+1$ prepare messages, it broadcasts the commit message. After the peer 
receives $2f+1$ commit messages, it locally confirms the transaction.
Since the peer has $2f+1$ commit messages, at least one honest voted for the transaction and no other can receive $2f+1$ different commit messages. This is also true for the prepare messages.
%Since no more than $f$ peers are Byzantine, at least one honest peer voted for this transaction

A non-leader Byzantine peer may delay messages or send incorrect messages. However, if the fraction $f$ of Byzantine peers is small, the honest peers are guaranteed to receive sufficient number of correct massages and then commit. That is, the actions of non-leader Byzantine peer may only delay the consensus.
A Byzantine leader may temporarily block the consensus by sending different messages to different peers or not sending messages altogether. In either case, the honest peers discover the Byzantine leader and replace it by forcing a \emph{view change}.  PBFT is guaranteed to withstand up to $f < n/3$ Byzantine peers regardless of the message propagation delay. %\cite{pbft}

%Since every peer has to communicate to all other peers, the complexity of the algorithm is $O(n^2)$ messages per single consensus.

\ \\
\textbf{SBFT.} Leader election is similar to PBFT. 
The algorithm works in four rounds. 
% status
(i) The generating peer sends its transaction to the leader. 
% proposal
(ii) The leader sends the proposal message to the peers. 
% commit
(iii) Once a peer receives the proposal message, it commits the transaction and sends the commit message to all other peers. 
If a peer receives  $f+1$ valid commit messages, it confirms the transaction. A Byzantine leader may be able to prevent some or all peers from committing by either sending different proposal messages to different peers or not sending messages at all. However, the honest peers discover such behavior and elect the new leader.
% notify, fix it status
(iv) The peer that confirms the transaction sends a notification message about the confirmation.
At the end of the fourth round, if there are peers that have not confirmed the transaction and terminated, the new leader is elected and the algorithm is repeated.

Similar to PBFT, this algorithm relies on at least one honest peer confirming the transaction. However, it assumes that there is a bound on communication delay between honest peers. If a message is not received after a certain delay, it is guaranteed never to arrive. On the other hand, the algorithm has to delay to ascertain this lack of message receipt. In practice this may make SBFT slower. However, it has higher resilience threshold.  It can tolerate up to $f < n/2$ Byzantine peers.

% need a picture here!

\begin{figure}[!t]
\small
\begin{tabbing}
1234\=12345\=12345\=12345\=12345\=12345\=12345\=12345\=12345\=12345\=\kill
$\textbf{constants}$\\
\> $gsize$ // group size\\
\ \\
$\textbf{variables}$\\
\>  $freeGroups$ \ \ \ // list of all the groups currently idle \\
\>  $waitingTrans$ \ \ \ // transaction waiting queue \\
\>  $activeComs$ \ \ \ // active committees \\
\ \\
$\textbf{commands}$ \\
new transaction  $t$ generated $\longrightarrow$ \\
\> add $t(ts)$ to tail of $transWaiting$ \\
\> $evaluate()$\\
\ \\    
consensus committee $c(gl)$ done $\longrightarrow$ \ \ \ // $gl$  is the list of groups in the committee \\
\>   remove $c$ from $activeComs$ \\
\>   add $gl$ to $freeGroups$ \\
\>   $evaluate()$ \\
\ \\   
$\textbf{function}$ \\
$evaluate()$ \\
\> let $t(ts)$ be at head of $waitingTrans$ \ \ \ // $ts$ is security level \\
\> $\textbf{while}(ts*gsize \leq size(freeGroups)\ \textbf{and}\ \textbf{not}\ empty(waitingTrans))$\\
\>\>    remove $t$ from head of $waitingTrans$ \\
\>\>    form committee $c$ from first $ts$ groups in $freeGroups$\\
\>\>    add $c$ to $activeComs$\\
\>\>    remove $ts$ groups from $freeGroups$\\
\>\>    let $t(ts)$ be at head of $waitingTrans$\\
\>\>    run consensus in committee $c$\\
\end{tabbing}
\vspace{-8mm}
\caption{Composite BlockGuard.}\label{figBC}
\vspace{-3mm}
\end{figure}

\ \\
\textbf{PoW.} Adding a transaction to the blockchain requires solving a computationally intensive problem involving the data from the new transaction. The new transaction is then cryptographically linked to last transaction in the blockchain. This task is \emph{mining} the new transaction. Once mined, the integrity of the transaction is easily verified.

The generating peer broadcasts the transaction to the network. Once a peer receives a new transaction, it attempts to mine it. After some peer mines the transaction, it broadcasts the mined transaction. Once a peer receives a mined transaction, it verifies it, attaches it to the blockchain and starts mining a new transaction on top of it. Several peers may mine transactions concurrently. This is a \emph{fork} in the blockchain. A branch of a fork may be extended by addition transactions mined on top of the current block. The shorter branch is discarded.

PoW consensus works correctly provided that the computational power of honest peers exceeds that of Byzantine peers. If peers have the same computational power, PoW consensus tolerates up to $f < n/2$ Byzantine peers.

\section{The Adaptive Security Problem and Solutions}

\textbf{The Adaptive Security Problem.} Consider a sequence of transactions that arrive over time. They need to be recorded into the blockchain. Each transaction has a security level. This security level is fixed and a solution can  neither modify nor anticipate it. Consider a \emph{committee consensus algorithm} that, given a fixed number of peers and a transaction, allows peers to agree to this transaction in a fixed number of steps. This agreement is effected despite the actions of Byzantine peers so long as the number of them is less than a certain resiliency threshold $f$. A peer may be in at most one committee at a time. For a transaction at security level $i$, the committee consensus algorithm needs to contain more peers than at level $i-1$. For example, a consensus algorithm may contain $2^i$ peers, i.e. the consensus committee size may grow exponentially with security levels. In this case, the number of security levels is $\log n$.

{\em The \emph{Adaptive Security Problem} requires the % a?
solution, an adaptive security algorithm, to assign committees to the transactions such that each committee satisfies the transaction security level.}

A trivial solution assigns all peers to every transaction. In other words, every transaction is processed at the highest security level. However, such solution is inefficient as communication resources are wasted with large committee agreeing on a low security level transaction. Another solution forms committees of appropriate size for each transaction but processes them sequentially. This solution uses the resources efficiently. However, it has low throughput since transactions are processed sequentially.  
Hence, we are led to consider an adaptive security algorithm that selects appropriate size committees and processes transactions with as much parallelism as possible. We present two such algorithms: \emph{Composite Blockguard} and \emph{Dynamic Blockguard}. 

\ \\
\textbf{Common features of the security algorithms.} We first discuss the features that are common to both Composite Blockguard and Dynamic Blockguard.
There are two committee types: reference and consensus. \emph{Reference committee} schedules transactions for verification. We assume its existence and do not discuss its formation and maintenance. Several papers discuss reference committee maintenance~\cite{rapidchain,elastico}. 

Once the new transaction is generated, it is \emph{pending}. The reference committee maintains a queue of pending transactions. Transactions are processed in FIFO order. If appropriate consensus committee is available, the pending transaction at the head of the queue is removed and dispatched to the committee for processing. The transaction is then \emph{tentative}. If the committee approves the tentative transaction, it is added to the blockchain and becomes \emph{recorded}. 
%Since transaction processing is FIFO, transactions behind the first one wait until  it is dispatched.
Once a sufficient size committee is available, the transaction is dispatched and the next transaction is considered. If enough peers are available, multiple transactions are processed concurrently.

\begin{figure}[tb]
% mention that we are ignoring unconfirmed transactions at the end of the DAG, MN
\small
\begin{tabbing}
1234\=12345\=12345\=12345\=12345\=12345\=12345\=12345\=12345\=12345\=\kill
$\textbf{constants}$\\
\> $winSize$\ \ \ // window size \\
\> $secMult$\ \ \  // security level multiplier\\
\\
$\textbf{variables}$ \\
\> $bc$ // blockchain DAG \\
\> $waitingTrans$\ \ \  // transaction waiting queue \\
\> $freePeers$ \ \ \ // list of peers not in committees\\
\> $activeComs$ \ \ \ // active committees \\
\\
$\textbf{commands}$ \\
new transaction $t$ generated $\longrightarrow$ \\
\>  add $t$ to tail of $waitingTrans$ \\
\\
end of recording stage $\longrightarrow$ \\
\>let $t(ts)$ be at head of $waitingTrans$\ \ \ // $ts$ is security level\\
\> let $peerlist$ be the first $winSize$ peers of totally ordered $bc$ \\
\> $\textbf{while} (size(peerlist) > ts * secMult\  
    \textbf{and}\ \textbf{not}\ empty(waitingTrans))$\\
\>\>  form committee $c$ by randomly selecting \\
\>\>\>$ts*secMult$ unassigned peers of $peerlist$\\
\>\> add $c$ to $activeComs$ \\
\>\> start consensus stage \ \ \ // run consensus algorithms in all committees of $activeComs$\\
\\          
end of consensus stage $\longrightarrow$\\ 
\> start recording stage \ \ \ // each committee records
accepted transaction into $bc$ \\
\end{tabbing}
\vspace{-9mm}
\caption{Dynamic BlockGuard.}\label{figDB}
\vspace{-5mm}
\end{figure}

\ \\
\textbf{Composite Blockguard adaptive security algorithm.} The algorithm is shown in Figure~\ref{figBC}. In this algorithm, peers are divided into storage groups maintaining independent blockchains. The algorithm maintains a list of idle groups $freegroups$ and stores pending transactions in $waitingTrans$. Once a new transaction arrives or a consensus committee is done, Composite Blockguard finds appropriate number of available groups, forms a consensus committee to process the next transaction in $watingTrans$ and dispatches the transaction. If not enough idle groups are available, the pending transactions wait.

% shishir
\ \\
\noindent\textbf{Dynamic Blockguard adaptive security algorithm.} The algorithm is shown in Figure~\ref{figDB}. This algorithm has a single blockchain and thus a single recording group. A consensus committee is selected out of this group of peers. Multiple consensus committees may operate concurrently if their members do not intersect. This means that the committees have to concurrently write to the same blockchain. To ensure the integrity of the blockchain, the computation proceeds by alternating two stages: consensus stage and recording stage. In the \emph{consensus} stage, committees agree on blocks to be written to the blockchain. Every committee must reach consensus before any committee may proceed to the next stage. In the \emph{recording} stage, each committee broadcasts the transaction to the group maintaining the blockchain. That is, they broadcast it to the whole network.  Each written transaction is cryptographically linked to all the written transaction in the previous recording stage. This way, the resultant blockchain is a series-parallel graph. %
% add stuff here maybe
%

\ \\
\emph{Committee selection window} is the set of unique peers that published in the blockchain most recently. Committee peers are picked at random from the committee selection window. For example, to select $64$-peer committee, a $512$ selection window may be chosen. Then the individual $64$ members are selected from the $512$ members of the window.

% may need a picture as well.

\ \\
\textbf{Algorithm analysis and comparison.}
In both Composite and Dynamic Blockguard, changing the security of each level or adding security levels is relatively easy. In Dynamic, it is just a matter of adjusting Committee Selection Window size or committee sizes. In Composite, the security levels can be changed by modifying the number of groups being merged into a committee

Composite Blockguard is simpler to implement. Since the groups do not overlap, the parallelism is potentially greater. On the other hand, Dynamic Blockguard automatically prefers the most active and, potentially, more reliable peers. However, Composite Blockguard has fewer synchronization issues between parallel committees as they are writing their results in separate ledgers.  Composite Blockguard, though, has the added complexity of \emph{cross-shard transactions} where a transaction affects more than a single blockchain. Dynamic Blockguard does not have this complication since it uses just one blockchain. 

\begin{figure}[htbp]
\vspace{-2cm}
\hspace{-2cm}
\begin{tabular}{@{}cc@{}}
\subfloat[Composite Blockguard, throughput]{
   \includegraphics{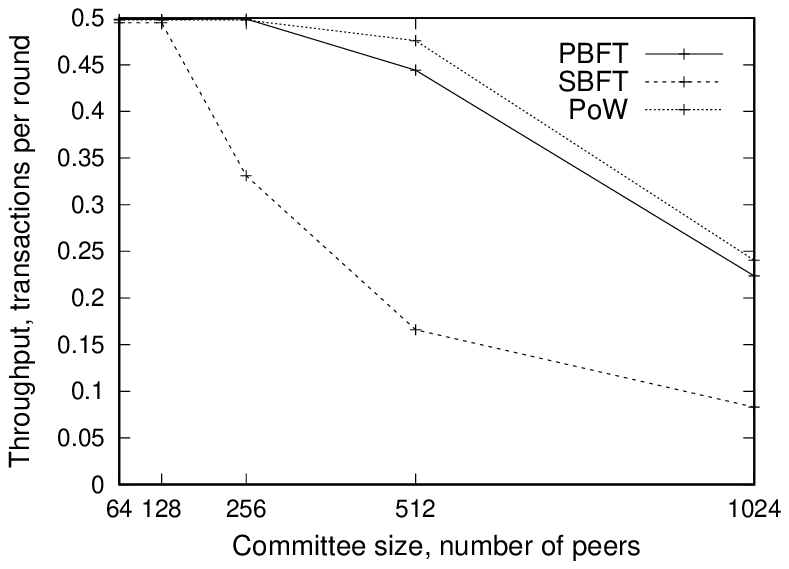}
   \label{figFixedThroughputComposite}
}&
% shishir replace this with your graph
\subfloat[Dynamic Blockguard, throughput]{
   \includegraphics{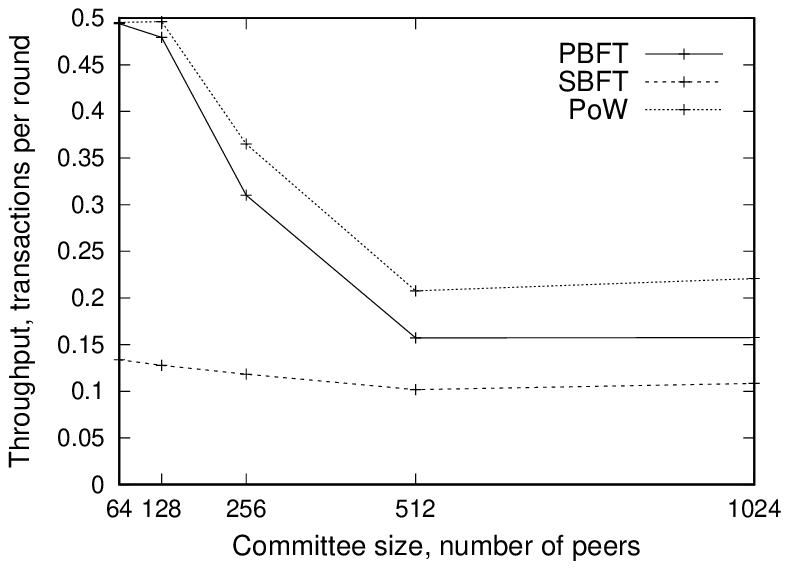}
   \label{figFixedThroughputDynamic}
}\\
\subfloat[Composite Blockguard, waiting time.]{
    \includegraphics{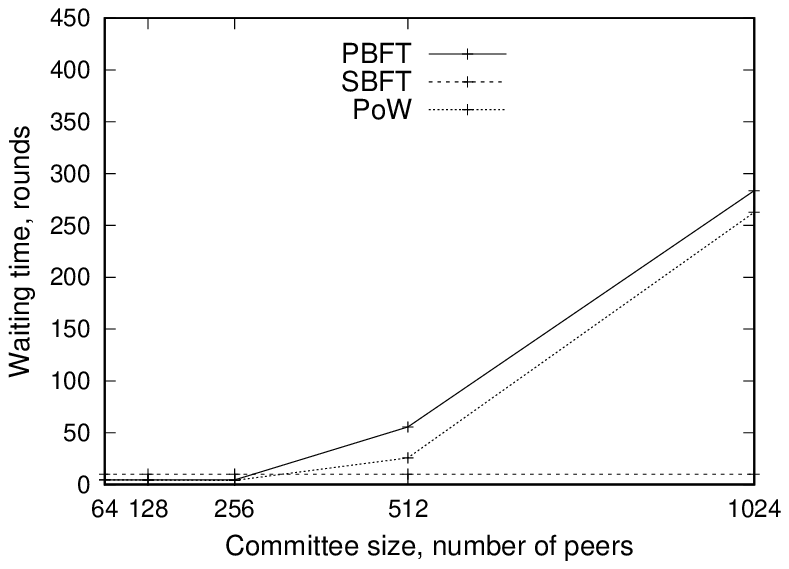}
    \label{figFixedWaitingTimeComposite}
}&
% shishir, replace this with your graph
\subfloat[Dynamic Blockguard, waiting time.]{
    \includegraphics{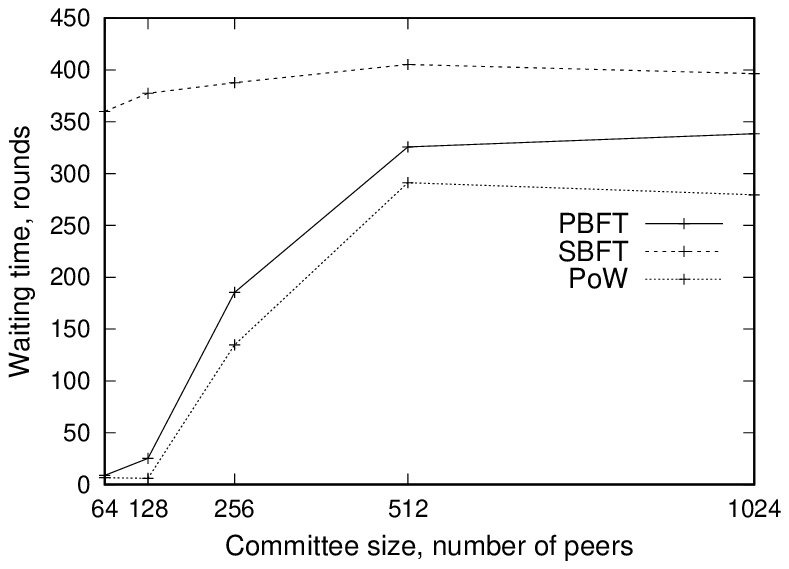}
    \label{figFixedWaitingTimeDynamic}
}\\
% kendric, place PBFT, fixed committees, varying ratio graph here
\subfloat[Composite Blockguard, PBFT, ratio of defeated committees]{
    \includegraphics{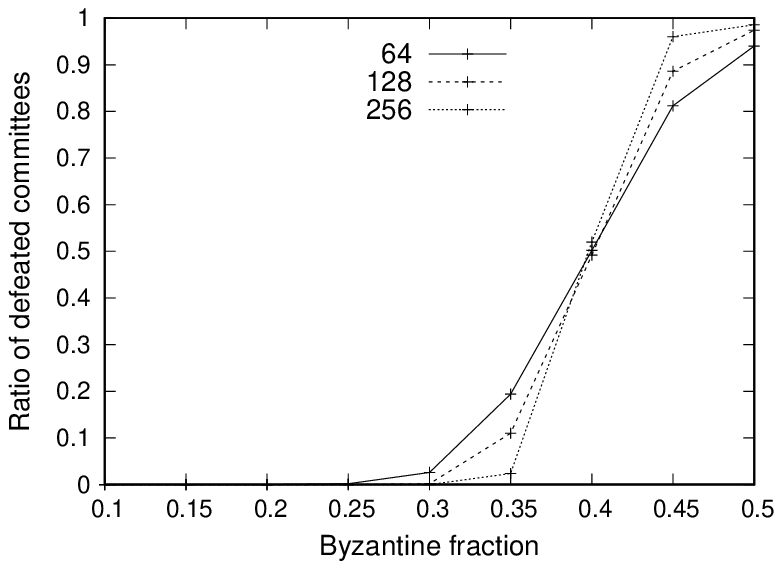}
    \label{figFixedDefeatedCommitteesComposite}
}&
% shishir, place PBFT, fixed committees, varying ratio graph here
\subfloat[Dynamic Blockguard, PBFT, ratio of defeated committees]{
    \includegraphics{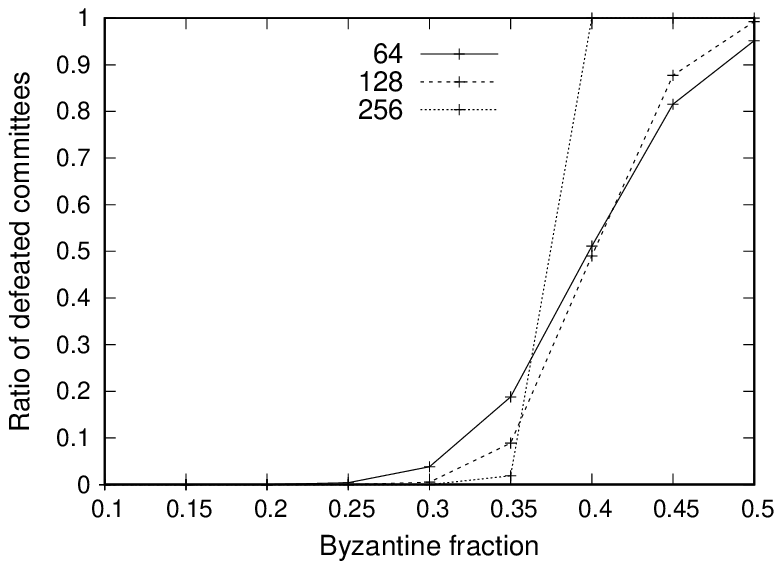}
    \label{figFixedDefeatedCommitteesDynamic}
}
\end{tabular}
\caption{Fixed size committees.}
\label{figFixed}
\end{figure}

\section{Performance Evaluation}

\subsection{Preliminaries}
\noindent\textbf{Setup.} We evaluate the performance of Composite and Dynamic Blockguard using abstract simulation. The code for our simulation is available on GitHub~\cite{github}.
The behavior of each algorithm is represented as a computation and the performance of such computations is evaluated.  
Individual computation consists of a sequence of rounds. In every round, each peer may receive a single new message, do local computation and send messages to other peers. 

Message propagation may take several rounds. The message propagation delay is uniformly distributed between one round and the maximum. If some peer $p_s$ sends several concurrent messages to the same peer $p_r$, message propagation delay is implemented as follows. Once $p_r$ receives the first message, the next message is delayed between one and the maximum number of rounds. Once this one is received by $p_r$, the next message delay is computed and so on. Message delivery is FIFO. In a single round the recipient may process only a single message from the same sender. However, if multiple messages from different senders are available for delivery by a single peer $p_r$, $p_r$ processes them all in a single round.

Unless specified otherwise, the fraction of Byzantine faults in the network is $n/10$. 
Byzantine behavior is difficult to simulate in its full complexity. We implement simplified Byzantine peer behavior as follows. A committee is \emph{reliable} if the number of Byzantine peers in it does not exceed its tolerance threshold. The committee is \emph{defeated} otherwise.
For example, a committee running PBFT is reliable if the number of Byzantine peers is less than $1/3$ of the total number of its members. The network may not recognize a defeated committee. A defeated committee proceeds operating as normal and writes its transaction to the blockchain. This transaction is counted as defeated. In PoW, if an honest peer mines a transaction in a defeated committee, the transaction is counted as reliable.

Byzantine peer behavior affects reliable committees as well. A committee running PBFT or SBFT may elect a Byzantine peer as a leader. In our simulation, an election of Byzantine leader forces a view change and, in effect, slows down the consensus. A correct transaction is, eventually, recorded in the blockchain.  Let us consider the operation of reliable PoW. If an honest peer mines a transaction first, it is recorded in the blockchain. If a Byzantine peer mines a transaction first, the transaction is discarded and mining re-starts. That is, similar to PBFT and SBFT, the presence of Byzantine peers slows down the operation of the consensus algorithm.  

In SBFT and PBFT, non-leader Byzantine peers have little influence over the performance of the algorithm, while Byzantine leaders are detected by the consensus algorithm. It is therefore possible that all Byzantine peers may eventually be detected and removed from the network. To counter this, our adversary may shuffle Byzantine peers. That is, a peer may start honest, become Byzantine and then become honest again over the course of our experiment. The adversary may never have more than $f$ Byzantine peers in the network at any given time. Let $s$ rounds be a shuffle period. Every $s$ rounds, a random number of Byzantine peers become honest and an equal number of honest peers become Byzantine. This maintains the ratio of Byzantine to honest peers in the network. Peers assigned to a committee may not be shuffled.  In Dynamic Blockguard, Byzantine peers are shuffled after the recording stage. In Composite Blockguard, shuffling happens every round but only affects non-assigned peers.

We use geometric distribution to select the security level of newly generated transaction. The selection probability is $50\%$. That is, for security level $k$, the probability of selection is $0.5^k$. For the highest security level, the probability is $0.5^{k-1}$. To put another way, half of the transactions are at the lowest security level $1$, then $25\%$ of transactions is at the next security level up and so on.

We do not take into account the performance of the reference committee in our experiments. However, we assume that the reference committee carries out the following tasks: it allocates peers to consensus committees, it conducts synchronous stages in Dynamic Blockguard.

\begin{figure}[!t]
\hspace{-2cm}
\begin{tabular}{@{}cc@{}}
\subfloat[Composite Blockguard, varying delay]{
    \includegraphics{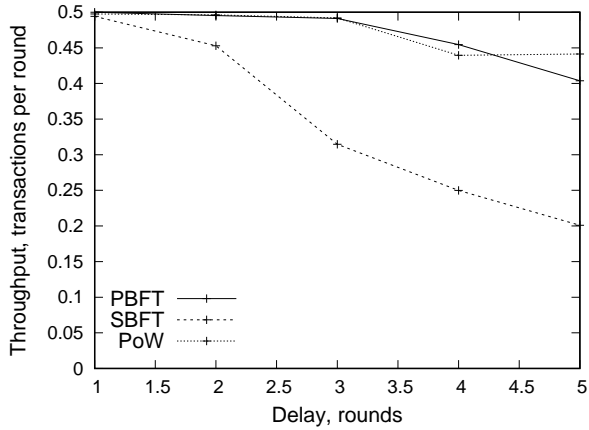}
    \label{figThroughputDelayComposite}
} &
\subfloat[Dynamic Blockguard, varying delay]{
    \includegraphics{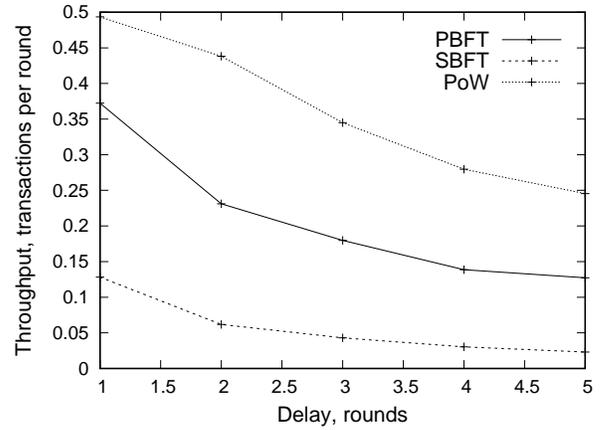}
    \label{figThroughputDelayDynamic}
} \\
\subfloat[Composite Blockguard, varying Byzantine fraction]{
    \includegraphics{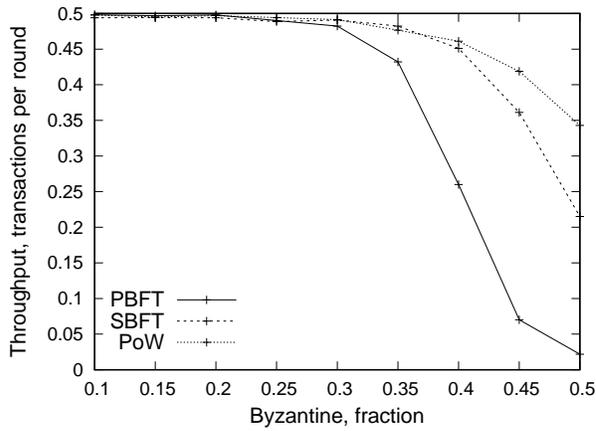}
    \label{figThroughputFractionComposite}
} &
\subfloat[Dynamic Blockguard, varying Byzantine fraction]{
    \includegraphics{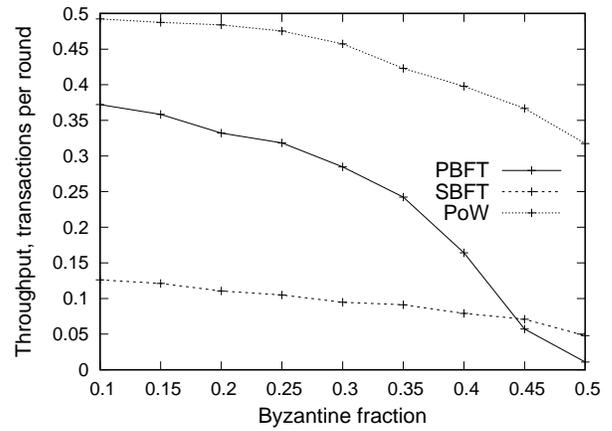}
    \label{figThroughputFractionDynamic}
} 
\end{tabular}
\caption{Throughput.}\label{figThroughput}
\vspace{-5mm}
\end{figure}

\ \\
\textbf{Experiment parameters and evaluation metrics.}
Unless stated otherwise, in the below experiments, the parameters are set as follows. The number of peers in the network is $1024$. The number rounds in a computation is $1000$. For each data point, we carry out $10$ computations and compute the average of the evaluated metric.

A new transaction is generated in every two rounds. 
% added this MN
This transaction generation rate slightly exceeds the maximum throughput of all consensus algorithms. 
The new transaction is generated by a randomly selected peer.

We have $5$ security levels. The highest security level is the 5-th level which contains the whole network. That is, the level-5 committee contains $1024$ peers. Each lower level contains half of the peers of the higher level. The lowest security level contains $64$ peers.

In PoW, we use binomial distribution to determine  the number of rounds it takes the peers to mine a transaction. The mode, i.e. most frequently occurring value, is $5$ and variance $2.5$.

% this sounds repetitive, MN
%
% The Composite does not have to synchronize transaction
% order between committees, as each group has its own unique blockchain. 

We vary maximum message delay and the fraction of Byzantine peers in the network. 
We compute the following metrics. \emph{Throughput} is the number of consensuses per round. We compute it as the number of successful consensuses divided by the length of the computation. Consensuses of defeated committees are not counted. \emph{(Transaction) waiting time} is computed as follows. For coordinated consensus algorithms, i.e. PBFT and SBFT, it is the number of rounds from the moment the transaction is generated till the first peer determines that the transaction is committed. For PoW, it is the time from this transaction. The waiting time for transactions of defeated committees is counted. 

\begin{figure}[!t]
\hspace{-2cm}
\begin{tabular}{@{}cc@{}}
\subfloat[Composite Blockguard, varying delay]{
   \includegraphics{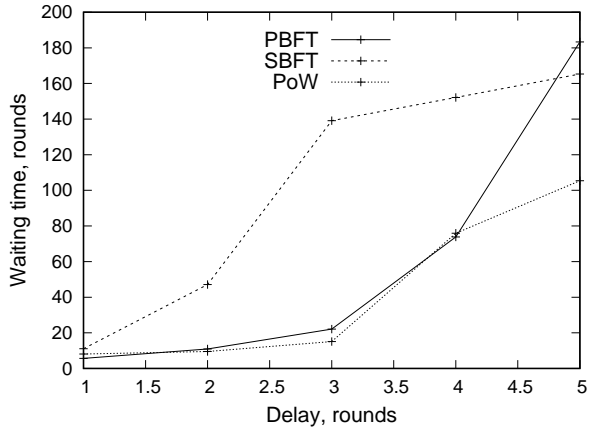}
    \label{figWaitDelayComposite}
}&
% shishir, replace this graph with yours
\subfloat[Dynamic Blockguard, varying delay]{
   \includegraphics{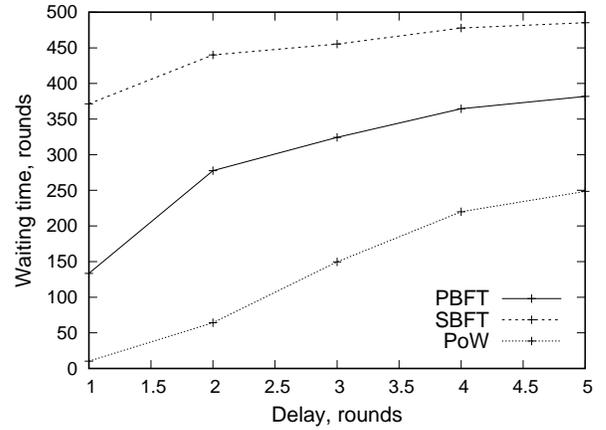}
    \label{figWaitDelayDynamic}
}\\
\subfloat[Composite Blockguard, varying Byzantine fraction]{
   \includegraphics{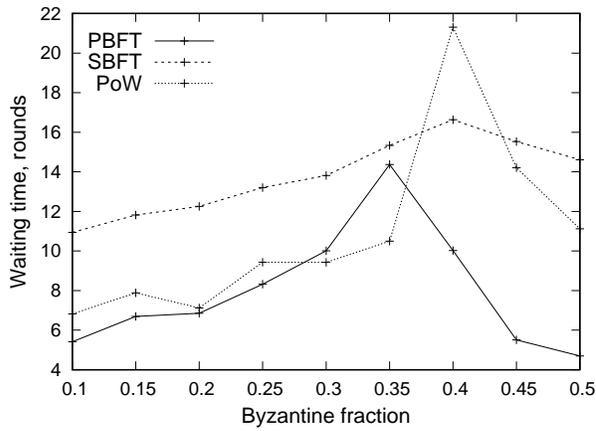}
    \label{figWaitFractionComposite}
} &
\subfloat[Dynamic Blockguard, varying Byzantine fraction]{
    \includegraphics{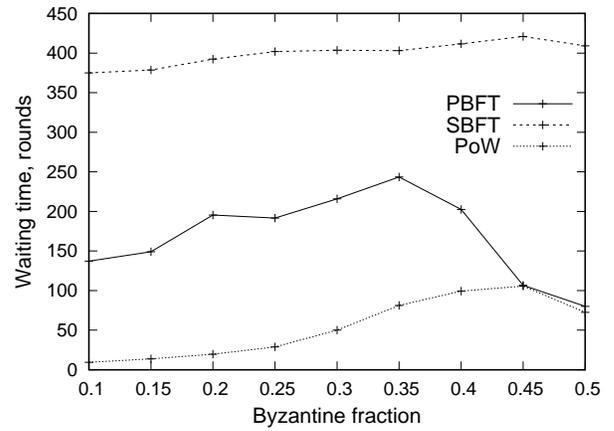}
    \label{figWaitFractionDynamic}
}
\end{tabular}
\caption{Average waiting time.}\label{figWaitingTime}
\vspace{-5mm}
\end{figure}

\subsection{Results and Analysis}

\textbf{Motivation experiments.} The results of the first series of experiments are shown in Figure~\ref{figFixed}. The results demonstrate the need for adaptive security. We show that there is a trade-off between the performance and the security of the network.  The security level of every transaction is ignored and all transactions are approved by committees of a specific size. We vary this single committee size. The committees proceed with maximum possible concurrency.

Figures~\ref{figFixedThroughputComposite} and \ref{figFixedThroughputDynamic} show throughput for Composite and Dynamic Blockguard respectively.
Similarly, Figures~\ref{figFixedWaitingTimeComposite} and~\ref{figFixedWaitingTimeDynamic} show waiting time for the two algorithms. The results indicate that as the committee size increases, the throughput declines and transaction waiting time increases.  The throughput decline is more pronounced for SBFT since this is a synchronous algorithm. It has to wait for the maximum delay time to ascertain the lack of message receipt. 
Similarly, wating time is greater for SBFT.

\begin{figure}[!t]
\hspace{-2cm}
\begin{tabular}{@{}cc@{}}

\subfloat[Composite Blockguard]{
    \includegraphics{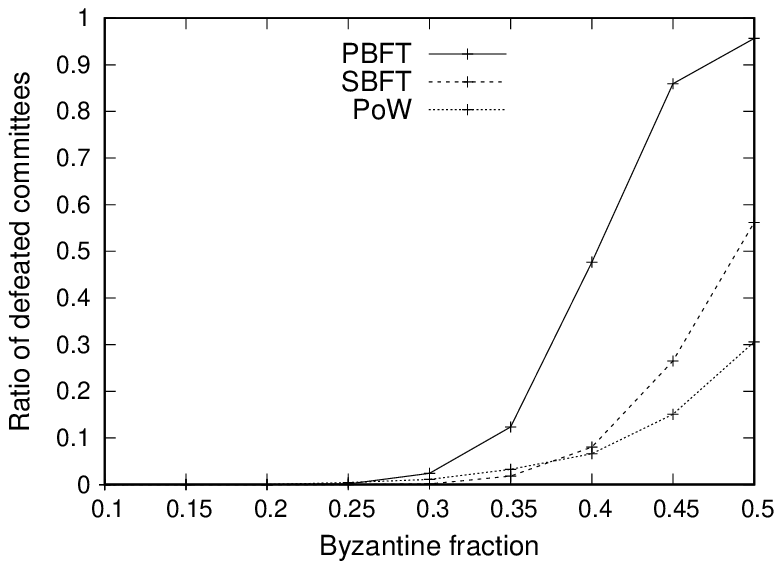}
    \label{figDefeatedComposite}
}&
\subfloat[Dynamic Blockguard]{
    \includegraphics{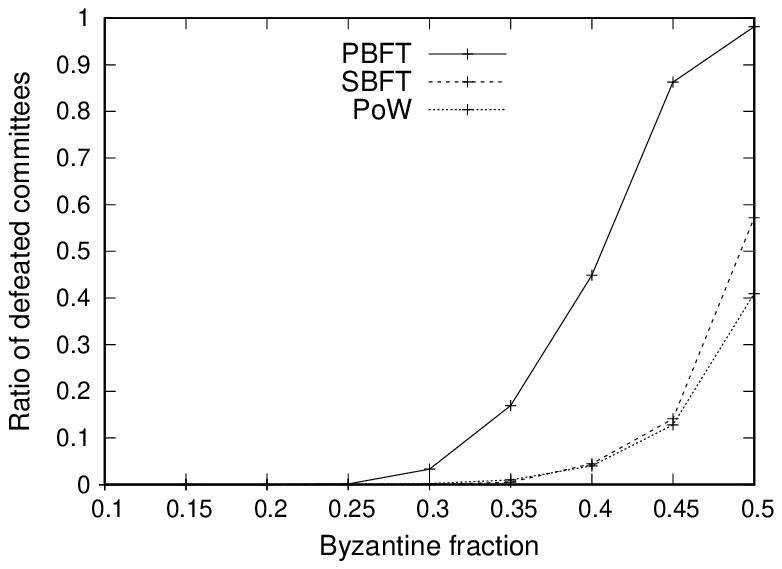}
    \label{figDefeatedDynamic}
}
\end{tabular}
\vspace{-3mm}
\caption{Ratio of defeated committees.} \label{figDefeated}
\vspace{-3mm}
\end{figure}

Conversely, Figures~\ref{figFixedDefeatedCommitteesComposite} and~\ref{figFixedDefeatedCommitteesDynamic} shows the ratio of defeated committees for particular fraction of Byzantine peers. We show the results for PBFT only and for the lower three security levels. The results for other consensus committee algorithms and security levels are similar. The results indicate that smaller size committees are defeated with greater ease.

\ \\
\textbf{Algorithm performance experiments.} The results of the performance evaluation of the adaptive security algorithms are shown in Figures~\ref{figThroughput}, \ref{figWaitingTime}, and~\ref{figDefeated}. Let us first discuss the results in Figure~\ref{figThroughput}. Figures~\ref{figThroughputDelayComposite} and~\ref{figThroughputDelayDynamic} demonstrate how throughput depends on the network delay for Composite and Dynamic Blockguard respectively. As network delay increases, the throughput declines. However, different consensus committees react to this increase differently. PBFT has the best performance and lowest decline since the committees just wait for the actual messages to arrive.  SBFT exhibits the most sensitivity to the network delay. The reason is that SBFT has to wait for the maximum delay to determine that the message is not coming.

Let us discuss Figures~\ref{figThroughputFractionComposite} and~\ref{figThroughputFractionDynamic}. It shows that the performance of Composite and Dynamic Blockguard decreases as the fraction of Byzantine peers in the network increase. This is due to Byzantine peers slowing down the consensus algorithms. 
PBFT suffers the most since its tolerance threshold is only a third of the peers. 

Let us address the results in Figure~\ref{figWaitingTime}.  Figures~\ref{figWaitDelayComposite} and~\ref{figWaitDelayDynamic} show the dependency of transaction waiting time on network delay. As expected, the waiting time increases with delay. SBFT is the most vulnerable to this increase since it has to wait for maximum delay time. Figures~\ref{figWaitFractionComposite} and~\ref{figWaitFractionDynamic} show how waiting time varies with the fraction of Byzantine peers. Let us explain the trends in the data. As the consensus committee approaches its resiliency threshold, the number of view changes or repeated transaction mining increases which increases the transaction waiting time. If the fraction is away from this threshold, the committees are either reliable or defeated. In either case the waiting time is relatively low. Thus, there is a peak near $n/3$ for PBFT and near $n/5$ for SBFT and PoW. This trend is less pronounced in Dynamic Blockguard since it is masked by synchronization across consensus committees in the same stage.

Let us now focus on the results in Figure~\ref{figDefeated}. The number of defeated committees increases with the fraction of Byzantine peers. It increases fastest for PBFT since it has the lowest tolerance threshold. It increases slowest for PoW since honest miners may still record a reliable transaction in a defeated committee. 

\ \\
% added this resume, MN
The results of our experiments indicate that both Composite and Dynamic blackguard algorithm provide adaptive security with a trade-off between performance and security parameters. Composite and Dynamic Blockguard  operate adequately regardless of the specific consensus algorithm that they use.

\section{Conclusions and Future Work}
In this paper, we defined the Adaptive Security Problem and showed two efficient solutions for it: Composite and Dynamic Blockguard algorithms. In conclusion, we would like to list further algorithm improvements and possible research directions. Composite and Dynamic Blockguard may be combined to further increase network efficiency. In both algorithms, rather than processing transactions FIFO, the reference committee may re-order transactions to better utlize available peers. Both algorithms have to be able to handle churn of peers. As old peers leave and new peers arrive, the algorithms have to be able to add them to the committees. To further demonstrate their practicality, both algorithms may be implemented and tested in a realistic blockchain system.

\bibliographystyle{plain}
\bibliography{blockguard}

\newpage
\section*{Appendix}
\subsection*{Parameter Experiments}
\begin{figure}
\hspace{-2cm}
\begin{tabular}{@{}cc@{}}
\subfloat[Composite Blockguard, throughput]{
   \includegraphics{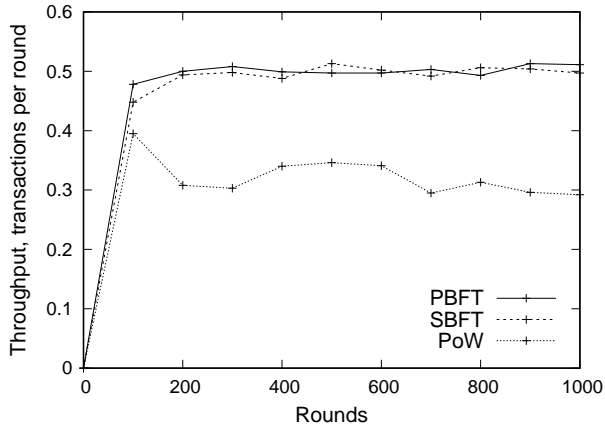}
   \label{figTimelineThroughputComposite}
}&
% shishir, below is Kendric's graph again, replace with yours
\subfloat[Dynamic Blockguard, throughput]{
   \includegraphics{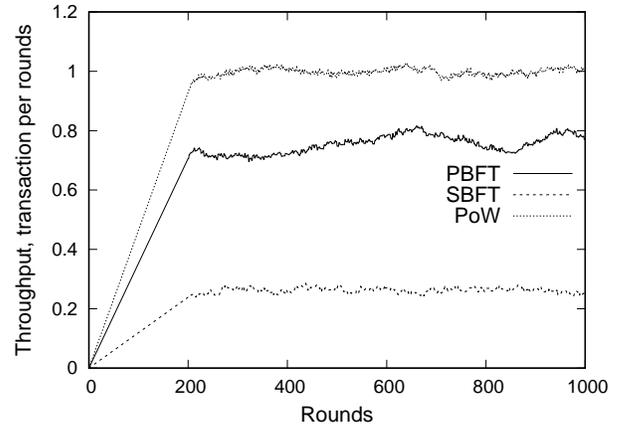}
   \label{figTimelineThroughputDynamic}
}
\\
\subfloat[Composite Blockguard, waiting time.]{
    \includegraphics{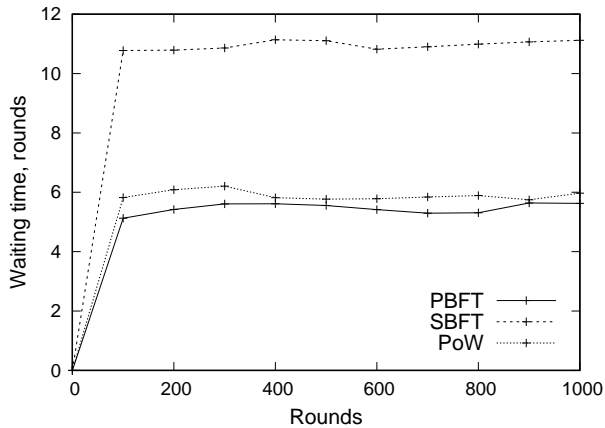}
    \label{figTimelineWaitingComposite}
}&
% shishir, below is Kendric's graph again, replace with yours
\subfloat[Dynamic Blockguard, waiting time.]{
    \includegraphics{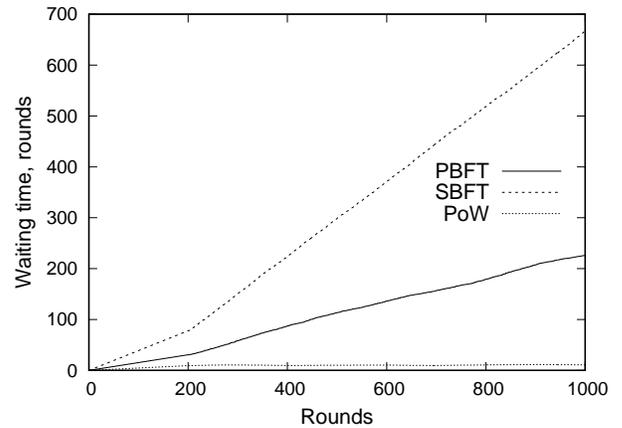}
    \label{figTimelineWaitingDynamic}
}
\end{tabular}
\caption{Timelines.}
\label{figTimeline}
\end{figure}
Let us discuss the results in Figure~\ref{figTimeline}. Each metric is taken as a rolling average where only transactions that have been confirmed in the last 200 rounds of computation are considered. For example, throughput is calculated by taking the total number of transactions over the past 200 rounds that have been confirmed over the total number of transactions submitted to the network in that time. In Figures~\ref{figTimelineThroughputComposite} and~\ref{figTimelineThroughputDynamic}, we see that while the throughput varies it is consistently confirming one transaction in two rounds for PBFT and PoW. In the case of SBFT, this throughput is lower: one transaction per 3 to 4 rounds. This is due to to the need for SBFT to wait for the maximum network delay. We note that  same trends in Figure~\ref{figTimelineWaitingComposite} for Composite Blockguard. In Figure~\ref{figTimelineWaitingDynamic}. the waiting time gradually increases for Dynamic Blockguard. This indicates that transactions are being submitted faster then the throughput of the network. This is not apparent in Figure~\ref{figTimelineThroughputDynamic} because throughput is measured as a rolling average. That is, the number of transactions confirmed by the network is constant but the queue of waiting transactions grows. The results show that Dynamic Blockguard can not handle as high of a transaction rate as Composite Blockguard. However, in order to compare the two, the same transaction rate should be used. The stability and consistency of the throughput shows that 1000 rounds is sufficient to take measurements on how the network will behave over a much longer duration. 
\end{document}